\documentclass[%
 reprint,
superscriptaddress,
 amsmath,amssymb,
 aps,
pra,
]{revtex4-2}

\usepackage{graphicx}
\usepackage{gensymb}
\usepackage{dcolumn}
\usepackage{bm}

\usepackage{comment}
\usepackage{hyperref}
\usepackage{multirow}
\usepackage{xcolor}
\usepackage{amsmath}
\usepackage{array}
\usepackage{longtable}

\newcommand{\der}[2]{\frac{\text{d} #1}{\text{d} #2}}

\newcommand{\simen}[1]{{\textcolor{black}{#1}}}
\newcommand{\new}[1]{{\leavevmode\color{black}#1}}

\begin{document}

\preprint{APS/123-QED}

\title{Understanding the temperature response of biological systems:\\
Part II - Network-level mechanisms and emergent dynamics}

\author{Simen Jacobs}
\thanks{These authors contributed equally.}
\affiliation{Laboratory of Dynamics in Biological Systems, Department of Cellular and Molecular Medicine, KU Leuven, Herestraat 49, 3000 Leuven, Belgium}

\author{Julian B. Voits}
\thanks{These authors contributed equally.}
\affiliation{Institute for Theoretical Physics, Heidelberg University, Philosophenweg 19, 69120 Heidelberg, Germany}
\affiliation{BioQuant-Center for Quantitative Biology, Heidelberg University, Im Neuenheimer Feld 267, 69120 Heidelberg, Germany}

\author{Nikita Frolov}
\thanks{These authors contributed equally.}
\affiliation{Laboratory of Dynamics in Biological Systems, Department of Cellular and Molecular Medicine, KU Leuven, Herestraat 49, 3000 Leuven, Belgium}

\author{Ulrich S. Schwarz}
\email{schwarz@thphys.uni-heidelberg.de}
\affiliation{Institute for Theoretical Physics, Heidelberg University, Philosophenweg 19, 69120 Heidelberg, Germany}
\affiliation{BioQuant-Center for Quantitative Biology, Heidelberg University, Im Neuenheimer Feld 267, 69120 Heidelberg, Germany}

\author{Lendert Gelens}
\email{lendert.gelens@kuleuven.be}
\affiliation{Laboratory of Dynamics in Biological Systems, Department of Cellular and Molecular Medicine, KU Leuven, Herestraat 49, 3000 Leuven, Belgium}

\date{\today}

\begin{abstract}
Building on the phenomenological and microscopic models reviewed in Part I, this second part focuses on network-level mechanisms that generate emergent temperature response curves. We review deterministic models in which temperature modulates the kinetics of coupled biochemical reactions, as well as stochastic frameworks, such as Markov chains, that capture more complex multi-step processes. These approaches show how Arrhenius-like temperature dependence at the level of individual reactions is transformed into non-Arrhenius scaling, thermal limits, and temperature compensation at the system level. Together, network-level models provide a mechanistic bridge between empirical temperature response curves and the molecular organization of biological systems, giving us predictive insights into robustness, perturbations, and evolutionary constraints.
\end{abstract}

\maketitle

\begin{figure}
\includegraphics[scale=0.9]{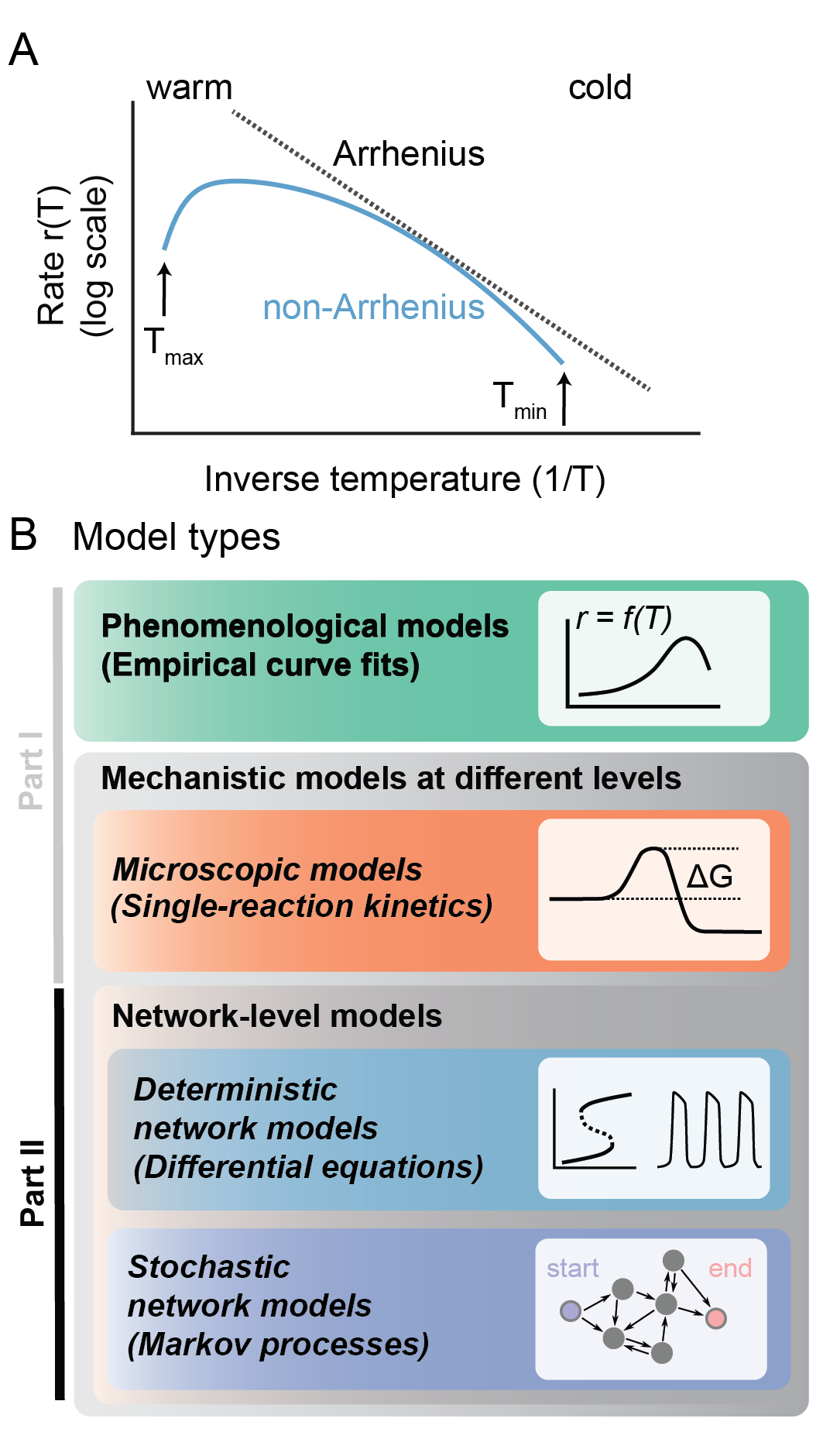}
\caption{\label{fig1}
\new{
\textbf{Temperature influences biological systems across scales and can be described using
different classes of models.} 
\textbf{(A)} Conceptual illustration of Arrhenius and non-Arrhenius behavior.  
In an Arrhenius plot (log rate versus $1/T$), simple reactions follow a straight line,
whereas biological processes typically show curvature and thermal limits ($T_{\min}, T_{\max}$).  
\textbf{(B)} Overview of modeling frameworks used to describe temperature responses, organized by
level of description.  
Phenomenological models provide empirical fits to observed rate–temperature curves, while
microscopic models derive rate–temperature relationships from reaction-level kinetics.
At a higher level, network-level models—either deterministic or stochastic—capture how
temperature affects coupled biochemical or regulatory systems.  
The phenomenological and microscopic approaches form the focus of Part~I \cite{Jacobs_review_pI}, whereas
deterministic and stochastic network models are the focus of this work in Part~II \cite{Jacobs_review_pII}.}
}
\end{figure}

\section*{Introduction}
In Part I of this review \cite{Jacobs_review_pI}, we surveyed phenomenological models that can describe biological temperature response curves, together with microscopic single-reaction level theories that derive temperature dependence from physical and chemical principles (Figure \ref{fig1}A-B). While these approaches clarify how temperature affects individual processes and observed rate curves, they do not capture how system-level temperature responses emerge from interactions among multiple reactions and regulatory interactions.

Here, in part II of this series, we focus on network-level mechanisms that generate emergent temperature dependence in biological systems (Figure \ref{fig1}B). In such models, temperature not only modulates individual reaction rates, but also the collective dynamics of interconnected pathways, feedback loops, and multistep processes. We consider two complementary classes of network-level descriptions. Deterministic models use coupled differential equations to describe how temperature shapes the dynamics of biochemical networks, while stochastic models (for instance formulated as Markov chains) capture the role of fluctuations, multi-step transitions, and statistical structure in determining system-level timing and rates.

Together, these network-level frameworks show how simple Arrhenius-like temperature dependence at the level of individual reactions can give rise to complex, non-Arrhenius temperature response curves at the system level (Figure \ref{fig1}A). By explicitly linking empirical temperature responses to network organization and dynamical structure, these models provide mechanistic insight into robustness, temperature compensation, and the constraints governing biological timing across scales.

\begin{figure*}[t!]
    \centering
    \includegraphics[scale=0.95]{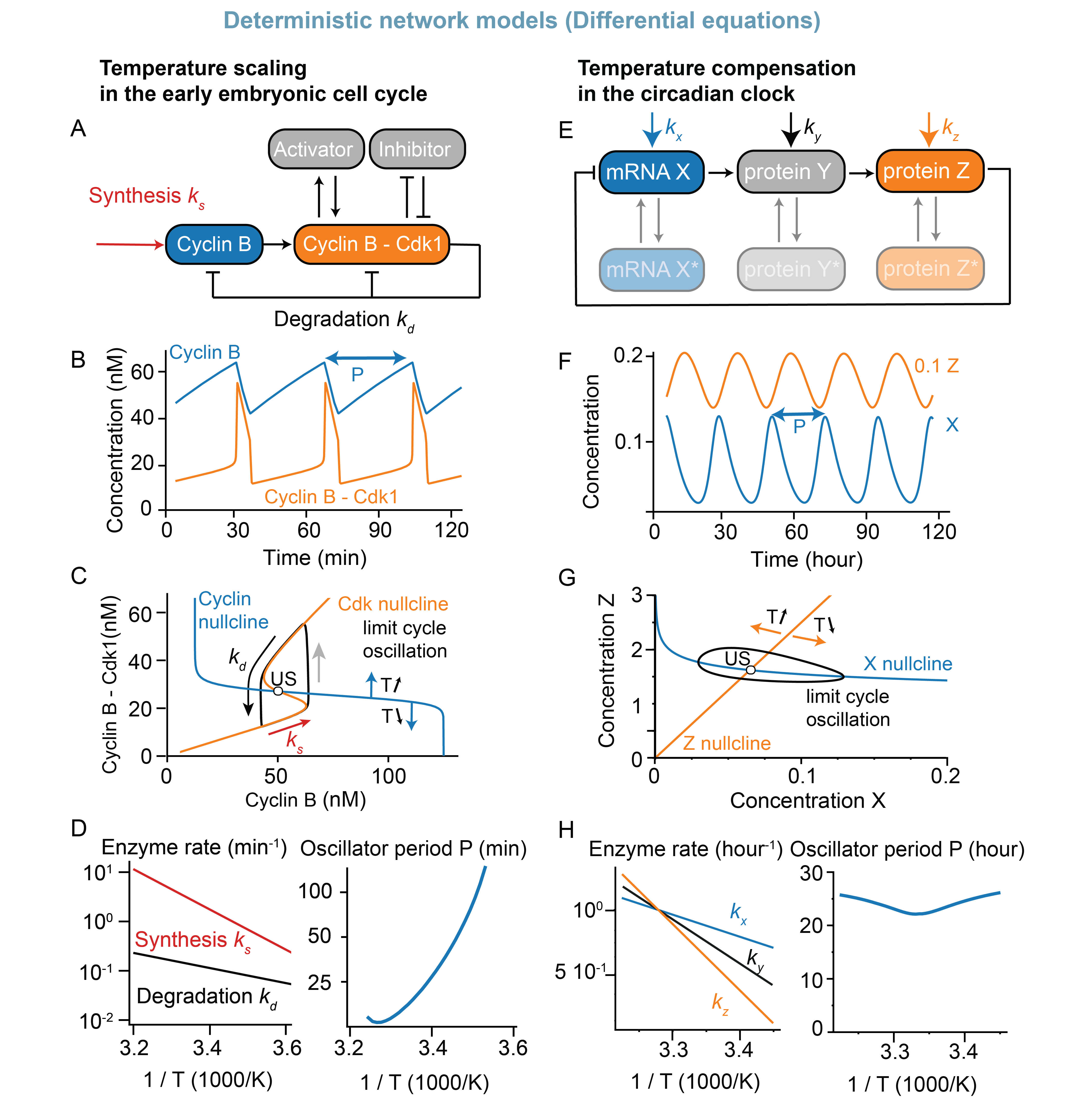}
    \caption{
    \textbf{Temperature response of biological oscillators.}
    \textit{A–D: Temperature scaling in the early embryonic cell cycle.}
    \textbf{(A)} Core regulatory architecture of the \textit{Xenopus} embryonic cell-cycle oscillator: Cyclin~B synthesis ($k_s$), Cdk1-dependent degradation ($k_d$), and a bistable activation module governing the switch-like transition between interphase and mitosis.
    \textbf{(B)} Characteristic sawtooth-like Cyclin accumulation and sharp Cdk1 activation pulses reproduced by the minimal two-ODE model of Yang \emph{et al.}\ \cite{Yang2013} (adapted from Rombouts \emph{et al.}~\cite{rombouts2025mechanistic}).
    \textbf{(C)} Phase plane of the two-variable model showing Cyclin and Cdk1 nullclines, the unstable steady state (US), and the emergent relaxation-type limit cycle.
    \textbf{(D)} Temperature scaling arises from differing activation energies of synthesis ($k_s$) and degradation ($k_d$): Arrhenius plots for the two rates (left) and the corresponding nonlinear, non-Arrhenius dependence of the model-predicted oscillation period (right). 
    \textit{E–H: Temperature compensation in a Goodwin-type circadian oscillator.}
    \textbf{(E)} Minimal transcription–translation feedback architecture underlying temperature-compensated circadian rhythms \cite{franccois2012adaptive,kidd2015temperature}. 
    \textbf{(F)} Example time series of X and Z generated by the Goodwin oscillator, showing robust 24-h rhythms.
    \textbf{(G)} Phase portrait of the Goodwin system, illustrating the X and Z nullclines, the unstable steady state (US), and the resulting limit cycle.
    \textbf{(H)} Temperature affects the synthesis rates ($k_x$, $k_y$, $k_z$) more strongly than degradation. Because the period is governed primarily by degradation, scaling only the production terms with temperature leaves the oscillation period nearly constant.
    }
    \label{fig:TempScalingOscillator}
\end{figure*}

\section*{Deterministic network models}

Mechanistic reaction–rate theories explain why particular functional forms arise at the level of
single biochemical steps, but they typically treat each reaction in isolation.  
By contrast, biological systems rely on regulatory networks, including feedback loops, switches, and
oscillatory modules. The temperature dependence of such a network cannot be inferred directly from the
properties of local reactions.  
This motivates \emph{dynamic network-based models}, in which temperature affects the
parameters of a coupled system of ordinary differential equations (ODEs), and emergent
properties arise from their collective nonlinear structure.

In deterministic network models, synthesis, degradation, activation, inhibition, and binding
rates are all explicit temperature-dependent parameters.  
Even when each individual step follows a simple Arrhenius or Eyring form, the nonlinear
interactions among components can produce strongly non-Arrhenius system-level behavior.
These models allow one to quantify how temperature reshapes nullclines, fixed points, and
bifurcation structure.  
For example, temperature-dependent changes in feedback strengths or reaction time scales can
shift Hopf bifurcations, alter oscillation periods, suppress or induce oscillatory regimes,
or generate partial temperature compensation.

Because these ODE models have a well-developed mathematical theory, they provide a natural
framework for studying how networks respond dynamically to temperature changes.  
Bifurcation analysis, eigenvalue spectra, slow–fast decompositions, and phase–plane geometry
offer mechanistic insight into how temperature alters stability, timing, and oscillatory
dynamics. Such features cannot be deduced from microscopic reaction kinetics alone.

Below, we illustrate this through two representative cases: temperature scaling in the early embryonic cell cycle oscillator, and temperature compensation in chemotaxis and circadian clocks.

\subsection*{Temperature scaling in an embryonic cell cycle oscillator}

A minimal and well-characterized example of a temperature-sensitive biochemical
oscillator is the early embryonic cell cycle of the frog \textit{Xenopus laevis}.
Entry into mitosis (M phase) and exit back into interphase are controlled by the
kinase Cdk1, which becomes active only when bound to Cyclin~B and appropriately
(de)phosphorylated. Active Cyclin~B--Cdk1 drives mitotic events and promotes
Cyclin~B degradation, completing a negative feedback loop.

The core regulatory architecture therefore consists of Cyclin~B synthesis, its
Cdk1-dependent degradation, and a bistable switch controlling Cdk1 activation
(Fig.~\ref{fig:TempScalingOscillator}A). Together, the slow accumulation and
degradation of Cyclin~B and the fast switching of Cdk1 form a classic relaxation
oscillator. In its simplest formulation, this system can be described by a
two-variable ODE model for Cyclin~B and active Cdk1
\cite{Yang2013,rombouts2025mechanistic}. The model reproduces the characteristic
sawtooth-like Cyclin~B dynamics and sharp Cdk1 activation pulses seen in extracts
(Fig.~\ref{fig:TempScalingOscillator}B). In the phase plane, the Cdk1 activation
module forms an S-shaped nullcline, and its intersection with the Cyclin~B
nullcline gives rise to an unstable state (US) around which a limit cycle
emerges (Fig.~\ref{fig:TempScalingOscillator}C).

To investigate temperature scaling, Arrhenius laws can be assigned to the
kinetic rates governing Cyclin~B synthesis ($k_s$), degradation ($k_d$), and Cdk1
activation and inactivation. Even this minimal two-ODE system is sufficient to
explain non-Arrhenius period scaling. When individual reactions have different
activation energies (Fig.~\ref{fig:TempScalingOscillator}D), the ratio
$k_s/k_d$ changes with temperature, shifting the Cyclin~B nullcline in the phase
plane. If $k_s$ increases more steeply with temperature than $k_d$, the Cyclin~B
nullcline moves upward. At high enough temperatures the nullclines cease to
intersect on the middle branch of the Cdk1 nullcline, the US disappears, and the
limit cycle collapses into a high-Cdk1 fixed point (persistent M phase).
Conversely, at low temperatures the system becomes trapped in a low-Cdk1
interphase-like state. These bifurcations generate thermal limits and a curved,
non-Arrhenius period–temperature relation (Fig.~\ref{fig:TempScalingOscillator}D).

Experimental measurements in cycling \textit{Xenopus} egg extracts support this
mechanistic picture \cite{rombouts2025mechanistic}. Within the viable embryonic temperature range, the rising
(interphase) and falling (M-phase) segments of the oscillation scale differently
with temperature, consistent with Cyclin~B
synthesis being more temperature-sensitive than its degradation. This difference
largely explains the non-Arrhenius scaling of the total period in the
physiological range. 


\subsection*{Temperature compensation in chemotaxis and circadian clocks}

The embryonic cell cycle oscillator illustrates how differences in activation energies among
reactions naturally generate non-Arrhenius scaling and thermal limits.  In contrast, some
biological oscillators maintain nearly constant period across a broad temperature range.
Understanding this temperature compensation requires analyzing how network structure
and feedback can counteract the intrinsic temperature sensitivities of individual reactions.

Temperature–robust signaling has been documented in several systems.  
In \textit{E.~coli} chemotaxis, receptors in different modified states respond oppositely to
temperature changes, and the antagonistic enzymes CheR and CheB display similar temperature
dependencies. Together with temperature–adjusted enzyme synthesis and degradation rates, these effects maintain
a stable signaling output across the physiological range \cite{oleksiuk2011thermal}.
While not an oscillator, this system illustrates the general principle that 
compensation can arise from counterbalancing temperature influences embedded in a regulatory network.

Circadian clocks provide another striking example because here the compensatory mechanisms must not just preserve steady signaling levels but also the period of a 24h rhythm.  
Temperature–independent circadian periods were first noted in classic work on the fiddler crab \cite{brown1948temperature} and the fruit fly \cite{pittendrigh1954temperature}. One noticed that daily rhythms in locomotor activity and color changes in the crab, and daily emergence of adult flies from pupae, persisted with almost constant period despite temperature variation.

Early mechanistic proposals by Hastings and Sweeney suggested that opposing biochemical reactions 
within the clock could have counterbalancing temperature responses \cite{hastings1957mechanism, forger2024biological}.  
This idea motivated a series of mathematical models in which different parts of the transcription–translation feedback loop scale differently with temperature 
\cite{leloup1997temperature,hong1997proposal,kurosawa2005temperature}.  
However, perfect compensation of temperature dependencies is not robust to 
parameter variation or genetic perturbations.

François, Despierre, and Siggia analyzed temperature compensation in the classic Goodwin-type transcription–translation oscillator \cite{franccois2012adaptive,kidd2015temperature}. The core architecture (Fig.~\ref{fig:TempScalingOscillator}E) consists of a negative-feedback loop in which a transcription factor drives production of an mRNA X, the mRNA is translated into a protein Y, and this protein can be further converted into a second form Z that feeds back to repress transcription of X. This minimal three-variable loop generates robust 24-hour oscillations (Fig.~\ref{fig:TempScalingOscillator}F–G), and a well-defined limit cycle in phase space. In this formulation, temperature primarily affects the production (synthesis) rates of X, Y, and Z, whereas the degradation rates play the dominant role in setting the oscillation timescale. Because the period depends almost exclusively on the degradation steps, the model exhibits a form of temperature compensation whenever the degradation rates are effectively temperature independent—even if synthesis accelerates with temperature. As shown in Fig.~\ref{fig:TempScalingOscillator}H, increasing temperature strongly affects the production rates, yet the period remains close to 24 hours.

However, degradation rates are not typically temperature-independent. To address this limitation, they proposed a more robust and biologically plausible mechanism based on adaptive buffering \cite{franccois2012adaptive,kidd2015temperature}. In their formulation, the molecular components of the oscillator do not exist in a single functional form but instead occupy multiple interconverting states, interpretable as distinct mRNA isoforms or post-translationally modified protein species. These interconversion reactions occur on a timescale comparable to temperature-sensitive synthesis and degradation. As temperature changes, the balance between these states shifts, and because each state contributes differently to the repression of transcription, this redistribution adjusts the effective feedback strength. In this way, the oscillator buffers the temperature dependence of the underlying biochemical rates and achieves robust temperature compensation without requiring fine-tuning of parameters.
Kidd, Young, and Siggia experimentally confirmed these predictions by measuring temperature responses in \textit{Drosophila} fly and mammalian systems \cite{kidd2015temperature}: perturbing the modification cycle disrupted compensation, whereas manipulating synthesis rates alone did not.

Recent work by Fu \textit{et al.} further generalized these results, showing that such adaptive modification cycles are not model-specific but represent a generic design principle of nonlinear oscillators with period-lengthening reactions \cite{fu2024temperature}. 


\new{
\subsection*{Influence of network structure}

The oscillator case studies above already show that temperature responses are shaped by network architecture, not only by the temperature dependence of individual reactions. The deterministic network wiring determines (i) which processes lie on the effective delay/slow manifold and (ii) which parameter combinations enter the period-setting feedback gain. A useful language for organizing these design principles is that of \emph{network motifs}, which are recurrent small subgraphs such as negative feedback loops and feedforward loops that implement stereotyped dynamical functions in transcriptional and signaling networks \cite{shenorr2002networkmotifs,milo2002networkmotifs,alon2007networkmotifs,mangan2003feedforwardloop}. Throughout this subsection we use temperature compensation to mean near-invariance of the oscillation period, and temperature robustness more broadly to include stability of other dynamical features (e.g., amplitude, transient waveforms, or noise sensitivity).

In oscillatory systems, motifs rarely appear in isolation: biologically realistic clocks often combine a delayed negative feedback backbone with additional positive feedback and/or post-translational modification cycles that reshape effective nonlinearity and delay \cite{tsai2008robustoscillations,chakravarty2023systematic}. In particular, a recent systematic comparison of minimal circadian oscillator motifs showed that a pure negative feedback (Goodwin-type) architecture (Fig.~\ref{fig:TempScalingOscillator}E) tends to yield the best temperature compensation of the period, whereas introducing positive feedback can improve robustness of the oscillation period to extrinsic noise. Moreover, combining positive and negative feedback can reduce the temperature dependence of noise sensitivity compared to either loop alone \cite{chakravarty2023systematic}.

This motif-level tradeoff links naturally to classic control-theoretic perspectives: (i) temperature compensation can emerge from balancing period-increasing and period-decreasing processes (``antagonistic balance'') \cite{ruoff1997modeling}, and (ii) motifs that implement (approximate) integral control can generate adaptation/insensitivity properties that resemble compensation in the appropriate input/output mapping \cite{ni2009controlcontroller}. Complementing these theoretical views, measurements of canonical motifs show that temperature robustness can depend strongly on operating regime: for example, strengthening a negative feedback loop can reduce the temperature sensitivity of transient timing, but at a performance cost (lower expression/dynamic range), and coherent feedforward loop behavior exhibits analogous robustness-performance tradeoffs \cite{patel2020temperaturemotifs}.

In circadian clock models for plants (which are typically multi-loop transcriptional networks), the scanning of broad parameter sets has led to the finding that temperature-robust solutions are enriched for strongly repressive interactions, autoregulation, and three-node feedback loops \cite{avello2021arabidopsis}. This complements motif-by-motif oscillator comparisons by showing that robustness  can be associated with specific wiring patterns across large regions of parameter space.

Finally, even when two oscillators share a similar coarse-grained architecture, where time delays and feedback exist in the network can alter the temperature-response outcomes. Both the embryonic cell cycle oscillator (Fig.~\ref{fig:TempScalingOscillator}A-D) and circadian models (Fig.~\ref{fig:TempScalingOscillator}E-H)  have a similar network architecture, combining negative feedback with additional regulatory steps that change the effective feedback strength \cite{Rombouts2023UpsDowns}. However, the consequences for temperature responses differ. In the circadian model, the interconversion between protein states occurs on timescales comparable to the temperature-sensitive synthesis and degradation reactions, allowing redistribution between isoforms to buffer temperature-induced changes and stabilize the period \cite{hong2007proposal}. In the embryonic cell cycle oscillator, by contrast, the dominant slow process is Cyclin~B accumulation, whereas the switch between Cdk1 states is very fast. This strong timescale separation means that changes in the relative temperature sensitivities of synthesis and degradation directly shift the nullclines and hence the period, decreasing the internal buffering. Thus, despite similarities in network structure, differences in relative timescales and in how feedback is routed lead to very different temperature responses.}

\begin{figure*}[t!]
    \centering
    \includegraphics[scale=1]{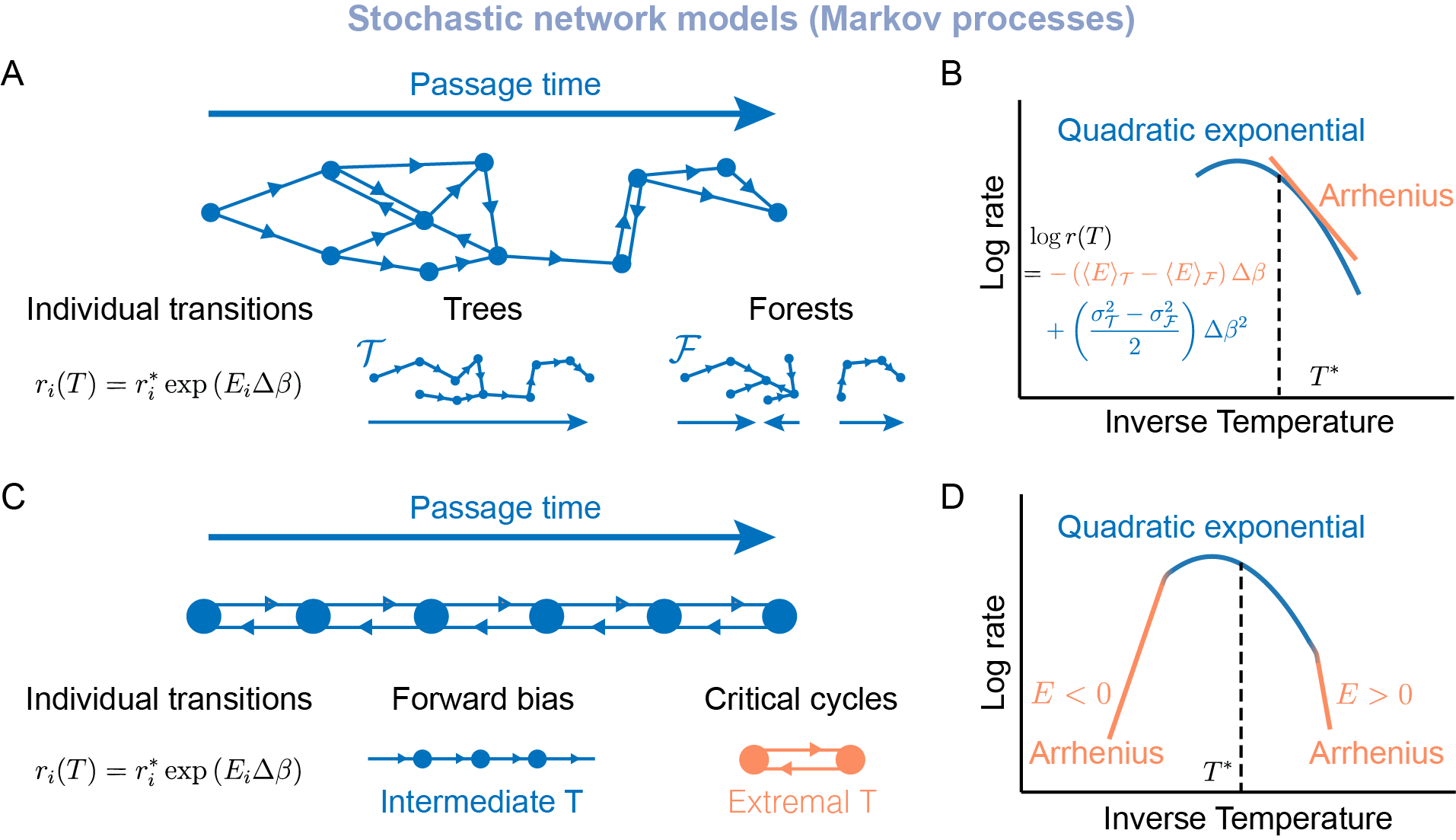}
 \caption{\textbf{Temperature response of stochastic network models.}
    \textbf{(A)} Schematic representation of a generic stochastic biochemical network, where individual transitions follow Arrhenius temperature dependence. Mean first-passage times can be expressed in terms of spanning trees ($\mathcal{T}$) and spanning forests ($\mathcal{F}$), whose total activation energies determine the cumulants entering the Taylor expansion of $\ln r(T)$ \cite{voits2025generic}. 
    \textbf{(B)} In large networks, the distributions of $E_{\mathcal{T}}$ and $E_{\mathcal{F}}$ become approximately Gaussian, causing all cumulants of order $n\geq 3$ to vanish and yielding a quadratic exponential dependence of the log-rate on inverse temperature.
    \textbf{(C)} A linear cascade of reversible reactions provides an analytically tractable example illustrating the emergence and breakdown of the quadratic exponential \cite{jacobs2025beyond}. Near the reference temperature $T^*$, forward-biased transitions dominate and collectively generate the generic quadratic form. At extreme temperatures, however, individual critical cycles control the mean first-passage time.
    \textbf{(D)} Resulting triphasic temperature response. Around $T^*$, the rate follows the quadratic exponential predicted for large networks, while below $T_+$ and above $T_-$ the rate reverts to Arrhenius behavior with positive and negative effective activation energies, respectively.
    }
    \label{fig:stochastic_networks}
\end{figure*}

\section*{Stochastic network models}
\simen{Stochastic network models represent biological processes as proceeding along a complex directed graph where the nodes represent states and the edges transitions. Assigning to each edge a temperature-dependent transition rate, this becomes a continuous time Markov-jump process. For a given process, one can then use the master equation to compute the distribution of first passage times between a given initial and final state \cite{van1992stochastic}. Most importantly, the first order moment of this distribution, the mean first passage time (MFPT), can then be equated to the inverse of the rate at which the process proceeds, while higher-order moments are usually expected to be minimized \cite{bel_simplicity_2009}. This allows to explicitly derive how temperature scaling laws emerge from stochastic network topology. }

\subsection*{Generic temperature response of networks}

Recent work has first established this statistical framework for the generic temperature dependence of large biochemical networks which are stochastic and for which each rate follows an Arrhenius law \cite{voits2025generic}. We note that even at the level of individual reactions, the Arrhenius form derived from Kramers’ theory relies on the assumption of sufficiently large energy barriers. When barriers become comparable to thermal fluctuations, this approximation breaks down and non-Arrhenius scaling can arise already at the single-reaction level, as shown using mean first-passage time approaches \cite{AbkenarPRE2017}. Starting from the master equation of a time-continuous, space-discrete Markov process, the analysis uses the graph-theoretic representation of the MFPT to a target state \cite{nam2023linear}. The overall rate $r$ is the inverse MFPT, and its temperature dependence $\ln{r(T)}$ can be expanded in a Taylor series whose coefficients are related to the distributions of the total activation energies along the spanning trees $E_\mathcal{T}$ and spanning forests of two trees $E_\mathcal{F}$ of the stochastic network (see Fig. \ref{fig:stochastic_networks} A) according to:
\begin{align} \label{eq:qaudr_exp_stoch_netw}
    \ln{r(T)}=\sum_{n=1}^\infty \frac{(-1)^n}{n!}(\kappa_n^{E_\mathcal{T}}-\kappa_n^{E_\mathcal{F}})\Delta\beta^n+const.,
\end{align}
where $\kappa_n^{E_\mathcal{T}}$ and $\kappa_n^{E_\mathcal{F}}$ denote the $n$-th cumulants of the distributions of $E_\mathcal{T}$ and $E_\mathcal{F}$, respectively. Here, $\Delta\beta:=\frac{1}{k_BT}-\frac{1}{k_BT^*}$, with $T^*$ being a reference temperature at which the activation energies and the prefactors of the Arrhenius equations can be treated as statistically independent. Biologically, $T^*$ corresponds to the temperature to which the organism has evolutionarily adapted.  

This result is valid for networks of any size and explains why biochemical networks often deviate from a simple linear Arrhenius relationship, as higher-order cumulants introduce systematic curvature in the Arrhenius plot. To predict the temperature response of a complex network, one has to characterize the total activation energy distributions of the spanning trees and spanning forests. Viewing the individual activation energies as independent random variables, the central limit theorem predicts that in the limit of large networks, the total activation energies approximate a normal distribution, as they are the sum of many of these individual activation energies. Then all cumulants of order $n\geq 3$ vanish, resulting in a quadratic shape in the Arrhenius plot (Fig. \ref{fig:stochastic_networks} B)
\begin{align}
    \ln{r(T)}=(\langle E\rangle _\mathcal{F}-\langle E\rangle _\mathcal{T})\Delta\beta +\frac{\sigma _\mathcal{T}^{2}-\sigma_\mathcal{F} ^{2}}{2}\Delta\beta ^2 +const,\label{eq:quadr_exp}
\end{align}
where $\langle E \rangle$ and $\sigma^2$ denote the means and variances of the respective energy distributions. This prediction agrees well with simulations of random networks and published developmental-rate data in flies which have previously been fitted by the phenomenological quadratic exponential described in Part I \cite{bliss1926temperature,powsner1935effects,crapse_evaluating_2021}. 
This limit requires that the difference in the activation energies along the trees and the forests remains distinct in the large network limit, typically assured by a bias in the activation energies for rates towards the target state. Moreover, it is only fully valid if the network is sufficiently complex to have a macroscopic number of spanning trees and forests. 

\subsection*{A linear cascade of reversible transitions}
The master equation approach has also been applied to a specific, and analytically tractable, network topology: a linear chain of $n$ reversible transitions (Fig. \ref{fig:stochastic_networks} C) \cite{jacobs2025beyond}. Such cascades provide a reasonable representation of many multi-step biochemical processes. However, since they have only one spanning tree they do not fulfill the requirements leading to the result in Eq.~\eqref{eq:quadr_exp} \cite{voits2025generic}. Still, the analysis of this model reveals that, even in this case, the generic quadratic exponential arises. Additionally, it shows how and why this quadratic scaling can break down at extreme temperatures.

Each individual step follows an Arrhenius law with a randomly distributed activation energy. At the reference temperature $T_0$, the cascade is assumed to be forward-biased, meaning that forward rates are, on average, larger than backward rates. Computing the MFPT, \simen{for example via the technique described in the previous section \cite{nam2023linear}},  from the initial to the final state shows that the overall temperature dependence of the rate falls into three regimes.

In the physiologically relevant range around $T_0$, the many forward transitions collectively dominate the MFPT. By the law of large numbers, the sum of their activation energies approaches a normal distribution. \simen{Just as in Eq.~\eqref{eq:quadr_exp} this leads to a quadratic exponential scaling for the overall rate
\begin{align}
    \ln r(T) = \langle E \rangle \Delta \beta + \frac{\sigma_E^2}{2} \Delta \beta^2 +  \ln \left(\frac{n}{\langle 1/r_f^*\rangle}\right),
\end{align}
where $E$ and $r_f^*$ denote the activation energies and forward rates of individual steps at $T^*$ and $\langle ... \rangle$ and $\sigma^2$ their mean and variance. }

At low temperatures, however, the system undergoes a sudden transition. As backward reactions become comparably faster, the system spends an increasing amount of time in cycles of forward and backward reactions before reaching the final state. Below a critical temperature $T_-$, a single cycle dominates the MFPT. In this regime, the rate scales Arrhenius-like with a large positive activation energy that depends on the sum and difference of the activation energies within this critical cycle.

A similar breakdown of the quadratic exponential occurs at high temperatures. Above a second critical temperature $T_+$, one cycle again dominates the MFPT leading to an overall rate that follows the Arrhenius law. However, in this regime the effective activation energy becomes negative. The critical temperatures separating the three regimes are:
\begin{align}
    \frac{1}{T_\pm} = \frac{1}{T^*} \pm \frac{R\sqrt{-\ln \left( \left \langle r_{f}^*/r_{b}^* \right \rangle \right) }}{\sigma_E},
\end{align}
where $r_{f}^*/r_{b}^*$ is the ratio between individual forward and backward reaction rates at the reference temperature. 

Together, these results show that the linear cascade exhibits an asymmetric triphasic temperature response: (i) a quadratic exponential near the reference temperature $T^0$; (ii) an Arrhenius regime with positive activation energy below $T_-$; and (iii) an Arrhenius regime with negative activation energy above $T_+$ (Fig. \ref{fig:stochastic_networks} D). This combined scaling law accurately captures the temperature dependence of a diverse set of biological processes when fitted to more than one hundred datasets that span species, traits, and timescales \cite{jacobs2025beyond}.

\section*{Discussion and conclusions}
Biological processes exhibit diverse temperature responses across scales, yet many of their large-scale features arise from shared physical and organizational principles. In Part I of this review \cite{Jacobs_review_pI}, we showed how phenomenological descriptions and microscopic reaction-level theories capture robust regularities in rate–temperature relationships and clarify why deviations from simple Arrhenius behavior are common. Here, in Part II \cite{Jacobs_review_pII}, we have focused on how these local temperature dependencies are transformed by network organization into emergent system-level temperature responses.

Network-level models demonstrate that temperature does not merely rescale individual reaction rates, but reshapes collective dynamics through feedback, branching pathways, timescale separation, and multistep processes. Even when each microscopic reaction follows a simple Arrhenius law, nonlinear interactions can generate curved Arrhenius plots, thermal limits, temperature compensation, or regime shifts. In this sense, non-Arrhenius system-level behavior is not an exception but a generic consequence of network structure and dynamical organization.

Deterministic and stochastic network descriptions provide complementary perspectives on this emergence. Deterministic models, formulated as systems of ordinary differential equations, offer mechanistic insight into how temperature alters phase–plane geometry, bifurcations, and oscillatory regimes. Stochastic models, by contrast, capture the statistical structure of multi-step processes and reveal generic scaling laws for timing and rates, particularly in systems where fluctuations and rare events dominate. Mean–first-passage–time frameworks show how universal temperature responses can arise from the aggregation of many Arrhenius-governed transitions, and why such scaling can break down at extreme temperatures. Several phenomenological observations that we emphasized in Part I \cite{Jacobs_review_pI}, such as quadratic scaling near the optimum and asymmetric thermal limits, emerge naturally from the network-level mechanisms reviewed here.

Together, the approaches reviewed in this two-part review series highlight that biological temperature responses reflect an interplay between local reaction physics and global network organization. A key direction for future work is to integrate measurements across scales, linking activation energies and molecular transitions to pathway topology, dynamical regimes, and physiological performance. Such multiscale approaches will be essential for understanding evolutionary adaptation and thermal robustness, predicting responses to fluctuating or warming environments, and ultimately for engineering temperature-sensitive or temperature-robust biological functions.

\subsection*{Data and code availability}
All original modeling code has been deposited at the Gelens Lab \textsc{Gitlab} [\url{https://gitlab.kuleuven.be/gelenslab/publications/temperature-review}], and is publicly available as of the date of publication. 

\subsection*{Declaration of generative AI and AI-assisted technologies in the manuscript preparation process}
During the preparation of this work the authors used ChatGPT in order to get feedback on grammar and phrasing. After using this tool, the authors reviewed and edited the content as needed and take full responsibility for the content of the published article.

\section*{Acknowledgements}
The work is supported by grants from Internal funds KU Leuven (C14/23/130, LG). JBV thanks the German Academic Scholarship Foundation (Studienstiftung des Deutschen Volkes) and USS the
Max Planck School Matter to Life, which is funded by the Dieter Schwarz Foundation and the Max Planck Society.


%

\simen{
\section*{Table of acronyms and technical terms}
\begin{table}[h]
    \centering
    \begin{ruledtabular}
    \begin{tabular}{cc}
     Term & Meaning \\ \hline
        Cdk1 & cyclin-dependent kinase 1 (protein) \\ 
        CheB & methylesterase (protein) \\
        CheR & methyltransferase (protein) \\ 
        M phase & mitosis \\
        MFPT & mean first-passage time \\
        ODE & ordinary differential equation \\ 
        US & unstable state \\
    \end{tabular}
    \end{ruledtabular}
    \caption{\simen{Alphabetical table of acronyms and technical terms used in the main text and their meaning.}}
    \label{tab:parameters_embryo_oscillator}
\end{table}
}

\section*{Mathematical details on the oscillator models}

\subsection{The embryonic cell cycle oscillator}

The embryonic cell cycle can be described by a minimal two-variable oscillator that tracks
the concentration of cyclin B and the activity of the Cyclin B –Cdk1 complex
(cdk1$_a$). Following the reduced formulation of \cite{rombouts2025mechanistic}, cyclin~B is
produced at a constant rate and degraded through APC/C when Cdk1 activity becomes
sufficiently high, while Cdk1 activation and inactivation are governed by fast
ultrasensitive feedbacks.

The dynamical equations are:
\begin{align}
    \frac{d\,\mathrm{cyc}}{dt} &= k_s - k_d\, d[\mathrm{cdk1}_a]\;\mathrm{cyc}, \\
    \epsilon\,\frac{d\,\mathrm{cdk1}_a}{dt} &=
        k_a\, a[\mathrm{cdk1}_a]\big(\mathrm{cyc} - \mathrm{cdk1}_a\big)
        - k_i\, i[\mathrm{cdk1}_a]\,\mathrm{cdk1}_a .
\end{align}

The ultrasensitive activation, inhibition, and APC/C–activation
functions are given by:
\begin{align}
    a[x] &= a_\mathrm{Cdc25} + 
        b_\mathrm{Cdc25}\,
        \frac{x^{n_\mathrm{Cdc25}}}{
              K_\mathrm{Cdc25}^{\,n_\mathrm{Cdc25}} + x^{n_\mathrm{Cdc25}}}, \\[4pt]
    i[x] &= a_\mathrm{Wee1} +
        b_\mathrm{Wee1}\,
        \frac{K_\mathrm{Wee1}^{\,n_\mathrm{Wee1}}}{
              K_\mathrm{Wee1}^{\,n_\mathrm{Wee1}} + x^{n_\mathrm{Wee1}}}, \\[4pt]
    d[x] &= a_\mathrm{APC} +
        b_\mathrm{APC}\,
        \frac{x^{n_\mathrm{APC}}}{
              K_\mathrm{APC}^{\,n_\mathrm{APC}} + x^{n_\mathrm{APC}}}.
\end{align}

For biologically realistic parameter choices, the system exhibits robust limit-cycle
oscillations with a period of $\sim$30 minutes, characteristic of early embryonic cleavage
cycles. In phase space, oscillations arise from slow Cyclin~B accumulation followed by a rapid
Cdk1 activation jump on the middle branch of the S-shaped Cdk1 nullcline, after which APC/C
activation drives a fast relaxation phase.

\vspace{0.5em}

Temperature dependence of the biochemical rates is introduced using Arrhenius scaling:
\begin{align}
    k(T) = k(T_0)\,\exp\!\left[-\frac{E_a}{R}
        \left(\frac{1}{T}-\frac{1}{T_0}\right)\right],
\end{align}
with reference temperature
\[
T_0 = 18^\circ\mathrm{C} = 291\ \mathrm{K}.
\]
Only the four core rate constants $k_s$, $k_d$, $k_a$, and $k_i$ are assigned nonzero
activation energies; all ultrasensitive functions are treated as temperature independent. The parameters for the example shown in Fig. 3 A-E are listed in Tab. \ref{tab:parameters_embryo_oscillator}.

\begin{table}[h]
    \centering
    \begin{ruledtabular}
    \begin{tabular}{ccc}
        Parameter & Value at $T_0$ & $E_a$ (kJ mol$^{-1}$) \\ \hline
        $k_s$  & $1.25\ \mathrm{nM\,min^{-1}}$ & 80 \\
        $k_d$  & $0.1\ \mathrm{min^{-1}}$       & 30 \\
        $k_a$  & $1\ \mathrm{min^{-1}}$        & 50 \\
        $k_i$  & $1\ \mathrm{min^{-1}}$        & 60 \\
        $\epsilon$ & $0.1$                     & 0 \\[6pt]
        $a_\mathrm{Cdc25}$ & 0.2 & 0 \\
        $b_\mathrm{Cdc25}$ & 0.8 & 0 \\
        $K_\mathrm{Cdc25}$ & $30\ \mathrm{nM}$ & 0 \\
        $n_\mathrm{Cdc25}$ & 10 & 0 \\[6pt]
        $a_\mathrm{Wee1}$  & 0.1 & 0 \\
        $b_\mathrm{Wee1}$  & 0.4 & 0 \\
        $K_\mathrm{Wee1}$  & $30\ \mathrm{nM}$ & 0 \\
        $n_\mathrm{Wee1}$  & 5   & 0 \\[6pt]
        $a_\mathrm{APC}$   & 0.1 & 0 \\
        $b_\mathrm{APC}$   & 0.9 & 0 \\
        $K_\mathrm{APC}$   & $30\ \mathrm{nM}$ & 0 \\
        $n_\mathrm{APC}$   & 15  & 0 \\
    \end{tabular}
    \end{ruledtabular}
    \caption{Parameters used for the embryonic cell cycle oscillator at the reference
    temperature $T_0 = 18^\circ$C. Only the core biochemical rate constants carry
    Arrhenius temperature dependence.}
    \label{tab:parameters_embryo_oscillator}
\end{table}

\subsection{The Goodwin-type circadian oscillator}

The Goodwin-type circadian oscillator shown in Fig. 3 E is governed by the following reaction equations:
\begin{align}
    \der{X}{t}&=k_x\frac{1}{1+Z^n}-d_XX,\\ \der{Y}{t}&=k_yX-d_YY,\\ \der{Z}{t}&=k_zY-d_zZ,
\end{align}
where $k_x,k_y,k_z$ and $d_x,d_y,d_z$ are the synthesis and degradation rates, respectively, and $n$ is a Hill coefficient with typical values $n=8-10$ for the Goodwin oscillator. Here, $n=9$ is used as in \cite{franccois2012adaptive}.

Introducing rescaled variables:
\begin{align}
    x&=\big(\frac{k_x}{k_y^nk_z^n}\big)^\frac{1}{n+1}X, \\y&=\big(\frac{k_xk_y}{k_z^n}\big)^\frac{1}{n+1}Y,\\ z&=\big(k_Xk_Yk_Z\big)^\frac{1}{n+1}Z,
\end{align}
in the limit $z\gg 1$, the dynamical equations read:
\begin{align}
    \der{x}{t}&=\frac{1}{z^n}-d_xx,\\ \der{y}{t}&=x-d_yy,\\ \der{z}{t}&=y-d_zz,
\end{align}
meaning that the oscillation period only depends on the degradation rates $d_x,d_y, d_z$, whereas the synthesis rates $k_x,k_y,k_z$ only affect the amplitude. 
So, if the degradation rates are temperature independent, one gets an oscillation with a robust period. 

As in the embryonic cell cycle model, temperature dependence is introduced through Arrhenius scaling of the synthesis rate constants:
\begin{align}
    k=k(T=T_0)e^{-\frac{E_a}{R}\big(\frac{1}{T}-\frac{1}{T_0}\big)},
\end{align}
where $T_0$ is a reference temperature at which the rates are specified.
The parameters for the example shown in Fig. 3 F-H are listed in Tab. \ref{tab:parameters_Goodwin_oscillator}.

\begin{table}[h]
    \begin{ruledtabular}
    \begin{tabular}{ccc}
         rate& value at $T_0$ & $E_a$ [kJ mol$^{-1}$] \\ \hline 
         $k_x$& $\kappa$ &20\\$k_y$& $\kappa$ & 40\\$k_z$& $\kappa$ & 60\\$d_x$& $0.2\kappa$ &0\\$d_y$& $0.2\kappa$ &0\\$d_z$& $0.2\kappa$ &0
    \end{tabular}
    \end{ruledtabular}
    \caption{The parameters used for the temperature compensated Goodwin-type model. The scale $\kappa=0.788$ $\text{h}^{-1}$ was chosen such that the period of one oscillation corresponds to a value of $24$ h. The reference temperature was set as $T_0=305$ K.}
    \label{tab:parameters_Goodwin_oscillator}
\end{table}

\end{document}